\documentclass[useAMS,usenatbib]{mn2e}
\usepackage{graphicx}
\title[New Galactic O2\,If*/WN6 star]{A Galactic O2\,If*/WN6 star possibly ejected from its birthplace
 in NGC3603}
\author[A. Roman-Lopes]{A. Roman-Lopes$^{1}$\thanks{E-mail:roman@dfuls.cl}\\
$^{1}$Physics Department - Universidad de La Serena - Av. Cisternas, 1200 - La Serena - Chile\\}

\begin{document}

\date{}

\pagerange{\pageref{firstpage}--\pageref{lastpage}} \pubyear{2010}

\maketitle

\label{firstpage}

\begin{abstract}
In this work we report the discovery of a new Galactic O2\,If*/WN6 star, a rare member of the extremely massive hydrogen core-burning group of stars
that due its high intrinsic luminosity (close to the Eddington limit), possess an emission-line spectrum at the
beginning of their main-sequence evolution, mimicking the spectral appearance of classical WR stars.
The new star is named WR42e and is found in isolation at 2.7\arcmin\ ($\sim$6 pc) from the core of the star-burst cluster NGC3603.
From the computed E(B-V) color excess and observed visual magnitude it was possible to estimate its absolute visual magnitude as M$_V$=-6.3 mag, which is a value 
similar to those obtained by other researchers for stars of similar spectral type both, in the Galaxy and in the Large Magellanic Cloud.
Considering the derived absolute visual magnitude, we computed a bolometric stellar luminosity of about 3.2$\times$10$^{6}$ L$_\odot$.
Finally, we estimate the mass of the new O2If*/WN6 star by comparing its observed magnitudes and colors 
with those of other probable NGC3603 cluster members, founding that the WR42e's initial mass possibly exceeds 100 M$_\odot$.

\end{abstract}

\begin{keywords}
 Stars: Wolf-Rayet; Infrared: Stars: Individual: WR42e;
Galaxy: open clusters and associations: individual: NGC3603
\end{keywords}

\section{ Introduction} 

NGC 3603 is one of the most massive
and dense star-forming regions known in the Galaxy,
being the closest exemplar of a star-burst like cluster (very similar to the star-burst cluster R136 in the Large Magellanic Cloud), 
with dozens of massive stars in its core, some of them possibly
presenting initial masses exceeding 100-150 M$_\odot$ \citep{b23}.\\
For a long time the O3 stars were considered to represent the most massive main-sequence stars. 
However, in the last decade it became clear that some hydrogen-rich nitrogen sequence
Wolf-Rayet stars, may be in reality extremely massive and luminous main-sequence stars \citep{b24,b25,b26,b23}, 
which due to their proximity to the Eddington limit presents an emission-line spectrum at the
beginning of their main-sequence evolution, mimicking the spectral appearance of classical WR stars.

From the empirical determination of stellar masses of high-mass binary systems, and the use of state of art stellar models it is becoming clear
that such kind of massive stars probably belong to the 
O2\,If*/WN6 and WNH spectral types \citep{b26,b10}, with this conclusion being supported by systematic studies of binaries made for
example by \citet{b25} for R145 (in 30 Dor), \citet{b26} for WR25, \citet{b28} for WR22 (both in the Carina 
Nebula), \citet{b29} for NGC 3603-A1, \citet{b9} and \citet{b8} for WR20a, \citet{b30} for WR21a and more recently \citet{b23} for some very massive exemplars 
in the center of the NGC3603 and R136 clusters.

In this letter we communicate the discovery of an O2\,If*/WN6 star (WR42e) at an angular distance of $\sim$ 2.7\arcmin\ (about 6 pc) from the NGC3603 cluster core, 
in apparent isolation in the field.
In section 2 we describe some aspects on the selection criteria, the observations and data reduction
procedures. In section 3 it is presented the results and some discussion, and finally in section
4 it is performed a summary of the work.

   \begin{figure*}
    \vspace{0pt}
   \centering
   \includegraphics[bb=14 14 704 486,width=9.0 cm,clip]{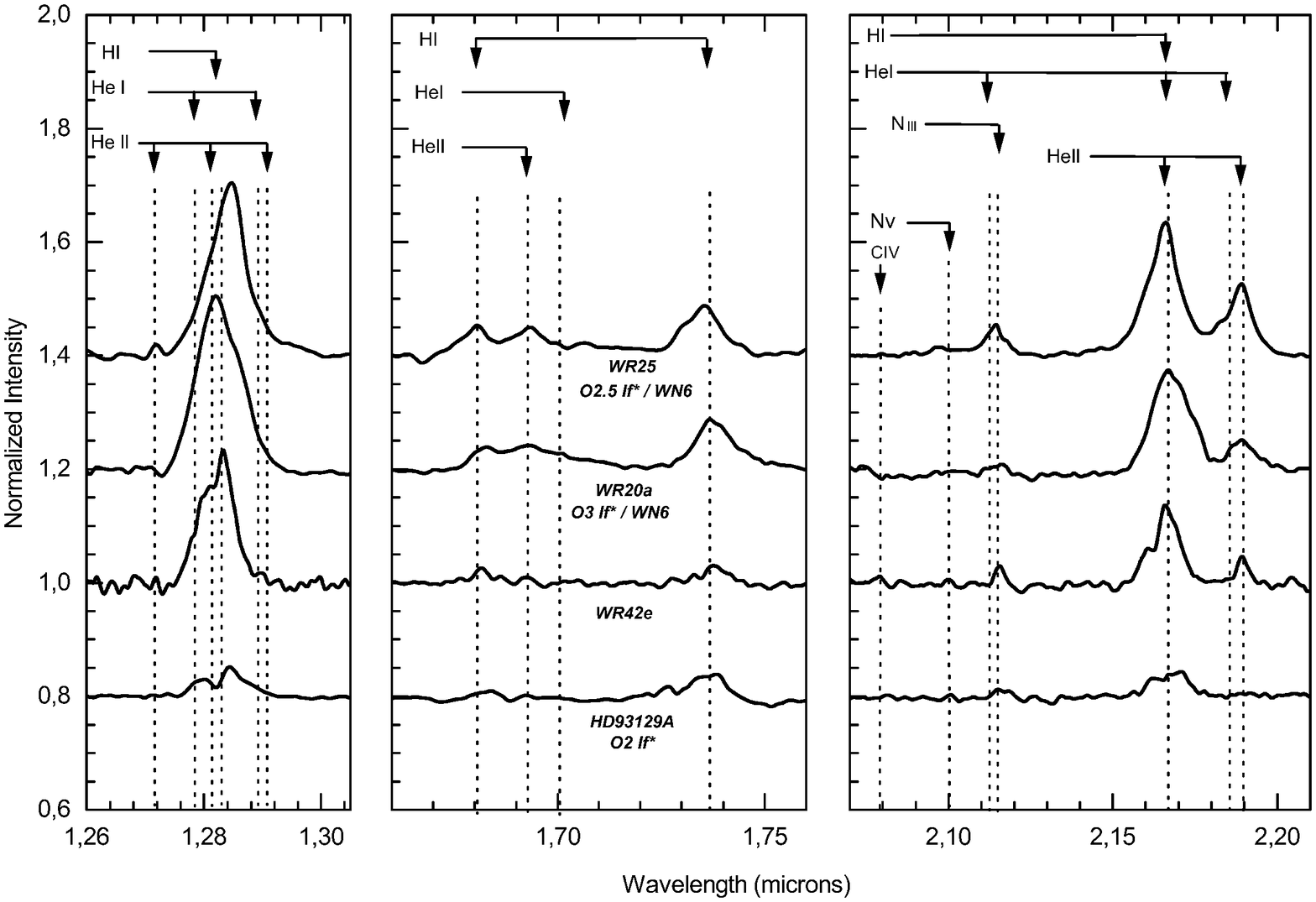}
      \caption{The J- H- and K-band continuum normalized SOAR-OSIRIS spectra of the star WR42e, together with the NIR spectrograms of
HD93129A (O2If*), WR20a (O3If*/WN6 + O3If*/WN6) and WR25 (O2.5If*/WN6).
The main H, He and N emission lines are identified by labels.}
         \label{FigVibStab}
   \end{figure*} 


\begin{table}
\caption{Summary of the SOAR/OSIRIS dataset used in this work.}
\label{catalog}
\centering
\renewcommand{\footnoterule}{}  
\begin{tabular}{cc}
\hline
   Date  & 09/05/2011\\
   Telescope  & SOAR\\
   Instrument & OSIRIS\\
   Mode  & XD\\
   Camera & f/3\\
   Slit  & 1$\arcsec$x 27$\arcsec$\\
   Resolution  & 1000\\
   Coverage ($\mu$m) & 1.25-2.35\\
   Seeing (\arcsec)  & 1-1.5\\
\hline
\end{tabular}
\end{table}

\begin{table*}
\begin{center}
\caption{Coordinates, Optical/NIR photometry, proper motion measurements (given in mas yr$^{-1}$), and X-ray and H$\alpha$ parameters of the newly-identified O2\,If*/WN6 star.
The UBVI photometry and the R-H$\alpha$ index were taken from \citet{b1}, while the Near-Infrared magnitudes are from \citet{b2}. The proper motion parameters were obtained
from the UCAC3 and PPMXL catalogs \citep{b18,b4} and the absorption-corrected 0.5-10keV flux is from the work of \citet{b3}.\label{tbl-2}}
\begin{tabular}{ccccc}
\\
\hline\hline
 RA = 11:14:45.50   &  Dec = -61:15:00.1 & U = 16.31 & B = 16.05 &    \\
\hline
V = 14.53 & I = 12.05 & J = 10.18$\pm{0.02}$ & H = 9.47$\pm{0.02}$ \\
\hline
K$_S$ = 9.04$\pm{0.02}$   &X-Ray (0.5-10 keV) = 3.3$\times10^{-17}$Wm$^{-2}$  &R-H$\alpha$ = 4.49 &pmRA (UCAC3) = -4.6$\pm{4.0}$   \\
\hline
pmDE (UCAC3) = 0.3$\pm{4.0}$ & pmRA (PPMXL) = -63.9$\pm{5.0}$& pmDE (PPMXL) = 1.4$\pm{5.0}$  &  \\
\hline\hline

\end{tabular}
\end{center}
\end{table*}

\begin{table*}
\caption{Wavelength and equivalent widths (\AA\ ) of the main emission lines detected in the SOAR-OSIRIS J-, H- and K-band spectra of WR42e.}
\label{catalog}
\begin{tabular}{ccccccccccc}
\hline\hline
& $\lambda$ ($\micron$)& 1.282& 1.681& 1.693& 1.736& 2.078& 2.100& 2.115& 2.166& 2.189 \\
\hline
& EW  ( \AA\ ) & -17.0$\pm2.5$& -1.5$\pm0.3$& $>$ -0.2& -2.3$\pm0.3$& $>$ -0.2& $>$ -0.2& -2.0$\pm0.3$& -20.2$\pm3.0$& -2.7$\pm0.6$  \\
\hline
\end{tabular}
\end{table*}

\section{Selection Criteria, Observations and data reduction}
In this section we comment on the selection criteria used to choose WR42e as target for Near-Infrared (NIR) spectroscopic follow-up, as well as the observations and data
reduction procedures.

\subsection{Selection criteria}

The newly identified O2If*/WN6 star (WR42e) was previously chosen for NIR spectroscopic follow-up observations based on its Near- to Mid-infrared colors, 
H$\alpha$ and X-Ray emission characteristics
and from its proximity with the NGC3603 star cluster. 
Details on the criteria and general selection methodology will be fully discussed in a forthcoming paper (Roman-Lopes in preparation).

\subsection{Observations and data reduction}

In this work we used NIR spectroscopic data acquired with the Ohio State Infrared Imager and Spectrometer (OSIRIS) at SOAR telescope. The observations were done
in 9th May 2011  with the night presenting good seeing conditions. Besides WR42e, we also acquired NIR spectra (to be 
used as comparison templates) for WR20a (O3If*/WN6 + O), WR25 (O2.5If*/WN6) and HD93129A (O2If*). 
In Table 1 it is presented a summary of the NIR observations.
The raw frames were reduced following standard NIR reduction procedures, which are presented in details in \citet{b5}, and shortly describe here.
The two-dimensional frames were sky-subtracted for each pair of
images taken at two nod positions, followed by division
of the resultant image by a master flat. 
The multiple exposures were combined, followed by one-dimensional extraction of the spectra.
Thereafter, wavelength calibration was applied using the sky lines, and
we estimate as $\sim$12$\AA$ the 1-$\sigma$ error for such calibrations, which was evaluate as one third of the Mean Full Width Half Maximum (MFWHM) of 
the sky lines detect in all three NIR bands.
Finally, the effects of the earth atmosphere in the science spectra were corrected using $J$, $H$ and $K$ band spectra of A type stars.

\section{Results}

\subsection{Optical and Near-Infrared photometry of WR42e}

Coordinates, photometric data and some other physical parameters of the new O2\,If*/WN6 star are shown in
Table 2. The U-, B-, V- and I-band magnitudes were taken from the work of \citet{b1}, while the near-IR photometry was obtained from the Two-Micron All Sky Survey
(2MASS, \citet{b2}). \\
Following \citet{b6,b7}, the star was named WR42e, and searching in 
SIMBAD\footnote{http://simbad.u-strasbg.fr/simbad/} 
we found that WR42e is identified as 2MASSJ11144550-6115001, being previously cataloged as source \#954 by \citet{b1}. It presents H$\alpha$ and X-Ray emission \citep{b1,b3},
and is not known to be a member of the NGC3603 cluster.

\subsection{NIR spectra of WR42e compared with those for HD93129A, WR20a and WR25}

In Figure 1 we present the telluric corrected (continuum normalized) J-, H- and
K-band spectra of WR42e and those for HD93129A (O2If*), WR20a (O3If*/WN6 + O) and WR25 (O2.5If*/WN6) \citep{b10}. 
The strongest features are the emission lines
close to the 1.282$\mu$m (Pa$\beta$) and 2.167$\mu$m (Br$\gamma$) hydrogen lines, as well as the He{\sc i}+N{\sc iii} and 
He{\sc ii} emission lines at 2.115$\mu$m and 2.189$\mu$m, respectively. An exception occurs in the case of prototype O2If* star HD93129A that 
(at least at the SOAR/OSIRIS spectral resolution and sensitivity) 
does not seen to present the HeII 2.189$\mu$m line in emission (or in absorption). Indeed this is the most evident difference (besides the relative intensities of the observed 
emission lines) between HD93129A spectrograms and those for the OIf*/WN6 stars. Interesting, the O2\,If* star R136a5 (30 Dor) does shows the 2.189$\mu$m line in emission, 
but in this case its intensity is too weak when compared with those found in OIf*/WN6 K-band spectrograms \citep{b10}.
In Table 3 it is shown the wavelength and corresponding equivalent widths (EWs) of the emission lines detected in the NIR spectra of WR42e.
From the comparison between the NIR spectra of the known O2If* and OIf*/WN6 stars with that for
WR42e, one can conclude that the it is probably an O2If*/WN6 star. 

   \begin{figure*}
    \vspace{0pt}
   \centering
   \includegraphics[bb=14 14 723 369,width=9.5 cm,clip]{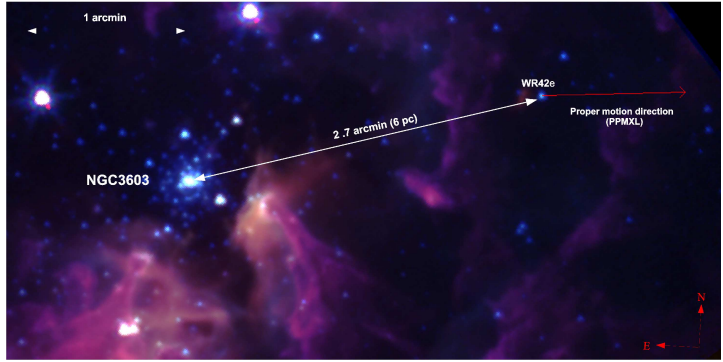}
      \caption{The Spitzer false-color images of the region towards NGC3603, with the cluster core and the new O2If*/WN6 star labeled. Also we represent the 
associated proper motion direction taken from the UCAC3 and PPMXL catalogs. North is to the top, east to the left.}
         \label{FigVibStab}
   \end{figure*}

\subsection{Spatial location and Interstellar reddening}

In Figure 2 it is shown a false color RGB image made from the 3.6$\mu$m (blue), 5.8$\mu$m (green) and 8$\mu$m (red) Spitzer IRAC images of the region towards NGC3603. 
There we label both, the core of NGC3603 and the newly found emission line star. As we can see, WR42e is found in relative isolation in the NGC3603 large field, being well 
displaced (2.7\arcmin) from the cluster core. O2If*/WN6 stars are very rare and normally expected to be found in the centre of the extremely massive clusters 
(like 30 Doradus and NGC3603 itself). In the specific case of NGC3603 all the presently known OIf*/WN6 and WN6ha (for example NGC3603-C and NGC3603-B) stars are found in its core, so 
one relevant question is: was WR42e formed in the place where it is now (in isolation), or was it formed inside the NGC3603 centre and then expelled
from its birthplace by its companions through some kind of dynamical interaction mechanism?
WR42e has a cluster core angular distance of 2.7\arcmin (about 6 pc), a value greater than the NGC3603 maximum cluster radius of 2\arcmin  \citep{b1}. However, such 
angular core distance is small when compared with those for other Galactic O2If*/WN6 stars found in isolation (e.g. WR20aa (16\arcmin) and WR20c (25\arcmin) - Westerlund 2 \citep{b12})
in the Galactic field. Indeed, the relative proximity to the NGC3603 core may be considered a strong indication that WR42e possibly was formed in the same birth site of the NGC3603 cluster.
On the other hand, the visual extinction to WR42e is another parameter that when compared with that for the NGC3603 cluster can give us additional clues about its possible
membership. We notice 
that the interstellar reddening law for the NGC3603 region is probably abnormal (\citet{b15,b1}), 
with a ratio of total to selective extinction value R$_V$=3.55$\pm$0.12 \citep{b1}. 
The mean E(B-V) color excess for the NGC3603 stellar population is in the 
range 1.25$\leq$E(B-V)$\leq$1.90, with the E(B-V) values increasing with the radial cluster center distance (\citet{b1}). From the Figure 5 of \citet{b1} one can see that 
for the NGC3603 cluster field, a cluster member candidate placed at a core distance of 2.7\arcmin, should present a (B-V) color excess in the range 1.65$\leq$E(B-V)$\leq$1.90. 
From Table 1 of the present work, we can see that
WR42e has (B-V) color $\sim$ 1.5 magnitudes, which for an assumed mean (B-V)$_0$ value of -0.3 magnitudes (typical for the hottest early-type stars), should corresponds 
to a color excess E(B-V) $\sim$ 1.8 magnitudes (A$_V$ $\sim$ 6.4 mag),  a reddening value fully compatible 
with what would be expected for the early-type cluster members placed at a cluster core angular distance in the range 2.0\arcmin-3.5\arcmin \citep{b1}. 

\subsection{Absolute magnitudes and mass of WR42e}

Once we have computed the color excess for the new star, it is possible to estimate its absolute magnitude using the distance modulus equation, 
assuming that the star is placed at an heliocentric distance (the same for NGC3603) of 7.6$\pm$0.4 kpc \citep{b23}. We computed M$_V$=-6.3 mag (or M$_K$=-6.1 
considering A$_K$=0.12A$_V$ - \citet{b23}), which are values similar to those obtained by other researchers for stars of similar type. As examples we can mention the
absolute magnitudes computed for WR20a (corrected for binary - O3If*/WN6 + O3If*/WN6), 
WR20aa and WR20c, (M$_V$ -6.3, -6.5 and -6.1 respectively) in Westerlund 2,
MK 30 (O2If*/WN5), MK 35 (O2If*/WN5) and R136a5 (O2If*) (M$_K$ -5.9, -6.0 and -6.0 respectively) in 30 dor, and HD93129A (O2If*) plus HD93131 (WN6ha) (M$_K$ -6.2 and 
-6.1 respectively) in the Carina region \citep{b23,b12,b10}. 
Considering the derived absolute visual magnitude and assuming a mean bolometric correction BC $\sim$ -4.2 mag \citep{b23,b10}, we estimate the bolometric magnitude of
the newly discovered star as M$_{Bol}$ $\sim$ -10.5 mags, which corresponds to a total stellar luminosity of 3.2$\times$10$^{6}$ L$_\odot$.
Also it is possible to estimate the mass of WR42e from its V- and I-band photometry. Indeed, we did that by comparing its observed magnitudes and colors 
with those of other probable NGC3603 cluster members, presented in Figure 7 of the work of \citet{b1}. 
In Figure 4 we present an adapted
version of their V $\times$ V-I diagram for NGC3603, with WR42e represented by a light blue circle. 
From this figure we can see that the initial mass of the new O2If*/WN6 star possibly exceeds 100 M$_\odot$.

   \begin{figure*}
    \vspace{0pt}
   \centering
   \includegraphics[bb=14 14 384 505,width=6 cm,clip]{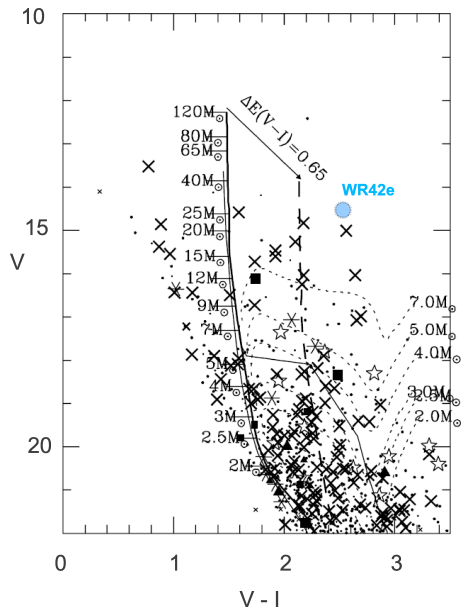}
     \caption{This is an adapted version of the V $\times$ V-I diagram (for NGC3603) of \citet{b1}, with WR42e represented by a light blue circle. 
From this diagram we can see that the current mass of the new O2If*/WN6 star possibly exceeds 100 M$_\odot$.}
         \label{FigVibStab}
   \end{figure*} 

Finally, WR42e was previously selected for spectroscopic follow up due to its associated X-ray emission (Chandra source 1BMC111445.5-611500).
From its absorption-corrected 0.5-10keV flux (Table 2), and considering an heliocentric distance of 7.6$\pm$0.4 kpc \citep{b23} 
we estimate its X-ray luminosity as L$_X$ $\sim$ 2.3$\times$10$^{32}$ erg s$^{-1}$, which compared to its bolometric luminosity results in 
L$_X$/L$_{Bol}$ = 5$\times$10$^{-8}$. This is a value compatible (considering the uncertainties in the measured X-ray flux, heliocentric distance, 
correction for absorption, etc) with what is expected ($\sim$ 10$^{-7}$) to be generated from single stars \citep{b31}.

\subsubsection{Could WR42e be a runaway stars?}

There is a reasonable consensus that very massive stars preferentially should be found in
the core of the most massive clusters, normally forming binary or multiple
systems.
However, it is an observational fact that a few very young and massive stars are
found in isolation in the field, being placed relatively far from any cluster or association of stars.
In this sense, there are two standard scenarios that try to account for such kind of objects.
One is known as \textit{dynamical ejection scenario} that involves dynamical
three- or multiple-body encounters in dense stellar systems, while the second is known as the \textit{binary-supernova scenario} which considers the
disruption of a short-period binary system from the asymmetric supernova
explosion of one of the binary components.
While the time scale for the binary-supernova scenario involves values larger than the
expected age (1-2 Myrs) for very massive hydrogen core burning stars \citep{b14}, the multiple ejection of massive stars from dense massive 
clusters in principle could account for the very young objects like 30 Dor \#16 in the LMC, and WR20aa and WR20c in the Milk Way.
For more details on the two models see for example the discussion in \citet{b21} and
references there-in. 

In the specific case of WR42e, we searched for proper motion data in order to
support the hypothesis of a runaway nature for this star, following the idea in which it was
ejected from its birthplace in the NGC3603 cluster center. In this sense, we did find proper motion measurements for WR42e in at least two catalogs. 
The most recent are those from
UCAC3 \citep{b18} and PPMXL \citep{b4} (see Table 2). In the later WR42 is identified as source 6101732644692891020, which presents a very high RA proper 
motion component 
of -63.9 mas yr$^{-1}$. On the other hand, from the UCAC3 catalog WR42e is identified as source 058-165111 that presents a quite moderate RA proper motion component value
of -4.6 mas yr$^{-1}$ (see Table 2). The origin of such discrepancy (within the quoted errors) between the RA proper motion measurements is unknown. 
As a comment, we should notice that the corresponding values for the DEC proper motion components of the two catalogs do not show such discrepancy.
In Figure 1 it is represented the corresponding proper motion vector (projected onto the sky plane) for WR42e. It may suggests that the star could be traveling in a 
trajectory opposite to that to the NGC3603 cluster center. 
However, while the first RA proper motion value (from the PPMXL catalogue) can be considered significant (taking into account the associated uncertainty), 
the second from the UCAC3 catalog is useless since the associated error is as large as the RA proper motion value itself.

Assuming that WR42e was formed in the NGC3603 core and using the proper motion values obtained from the PPMXL catalog, we can estimate 
the epoch the star was ejected from its birthplace in NGC3603. In addition, considering 7.6 kpc 
as the heliocentric distance to the NGC3603 cluster \citep{b23}, we can obtain estimates for the presumed WR42e's projected traveling velocity. 
In the case of the proper motion values given by the PPMXL catalog, one finds that the star
should be ejected from the NGC3603 center at only 2500 years ago, which for the quoted distance would imply in an
incredible traveling projected velocity of more than 2000 km s$^{-1}$, a value too high to be considered reliable. The reason for such discrepancy probably 
resides in a wrong value for the RA proper motion presented in the PPMXL catalog (perhaps due a typo error).
On the other hand, considering that the age of the most massive stars in NGC3603 has recently been derived to be 1.5$\pm$0.1 Myr \citep{b23}, this leaves 0.5 Myr
time for the possible dynamical interaction between the cluster members and the associated ejected O2If*/WN6 star \citep{b32}. In this sense,
we can estimate a lower value for the traveling velocity of WR42e assuming that the star was ejected 1 Myr ago from the center of
NGC3603. From the observed angular distance of $\sim$ 2.7\arcmin ($\sim$6 pc), a traveling time of 1 Myr would correspond to a proper motion of about 0.16 mas yr$^{-1}$,
which in turns results in a somewhat lower traveling velocity of 6 km s$^{-1}$. However, \citet{b21} argue that 
the average velocity attained by the 70-80M$_\odot$ stars can be quite moderate ($<$30 km s$^{-1}$), so that they cannot be formally classified as runaways.

\section{Summary}

In this work we report the discovery of a new Galactic O2\,If*/WN6 star, a rare member of the extremely massive hydrogen core-burning group of stars
that due its high intrinsic luminosity (close to the Eddington limit), possess an emission-line spectrum at the
beginning of their main-sequence evolution, mimicking the spectral appearance of classical WR stars.
The new star is found in isolation at 2.7\arcmin\ ($\sim$6 pc) from the core of the star-burst cluster NGC3603.
From the computed E(B-V) color excess and observed visual magnitude it was possible to estimate its absolute visual magnitude as M$_V$=-6.3 mag (or M$_K$=-6.1 
considering A$_K$=0.12A$_V$ - \citet{b23}), which are values similar to those obtained by other researchers for stars of similar spectral type both, in the Galaxy and in the
Large Magellanic Cloud.
Considering the derived absolute visual magnitude and assuming a mean bolometric correction BC $\sim$ -4.2 mag \citep{b23,b10}, we estimate the bolometric magnitude of
the newly discovered star as M$_{Bol}$ $\sim$ -10.5 mag that corresponds to a total stellar luminosity of about 3.2$\times$10$^{6}$ L$_\odot$.
Finally, we estimate the mass of the new O2If*/WN6 star, by comparing its V- and I-band photometry 
with those of other probable NGC3603 cluster members, founding that the WR42e's initial mass possibly exceeds 100 M$_\odot$.

\section*{Acknowledgments}

We would like to thank the anonymous referee by the careful reading of the manuscript. Her/his
comments were valuable to improve the clarity and presentation of this letter.
This work was partially supported by the Department of Physics of the Universidad de La Serena. 
This research has made use of the NASA/ IPAC Infrared Science Archive, which
is operated by the Jet Propulsion Laboratory, California Institute of
Technology, under contract with the National Aeronautics and Space
Administration. 
This publication makes use of data products from the Two Micron All Sky
Survey, which is a joint project of the University of Massachusetts and the
Infrared Processing and Analysis Center/California Institute of Technology,
funded by the National Aeronautics and Space Administration and the National
Science Foundation. 
This research made use of the SIMBAD database.  
ARL thanks financial support from Diretoria de Investigaci\'on - Universidad de La Serena through Project \textquotedblleft Convenio de desempe\~no DIULS CDI12\textquotedblright.
ARL thanks the staff of the SOAR Telescope by the help during the observing runs.

\end{document}